\documentclass[numbered]{trbunofficial}
\usepackage{graphicx}
\usepackage{xcolor}
\usepackage{diagbox}
\usepackage{longtable}
\usepackage{adjustbox}
\usepackage{booktabs}
\usepackage{yhmath}
\usepackage{mathtools}
\usepackage{mathrsfs} 
\newcolumntype{L}{>{$}l<{$}} 
\newcommand{\iidequal}{\overset{\mathrm{i.i.d.}}{=\joinrel=}}

\usepackage[hidelinks]{hyperref}
\usepackage{derivative} 

\AuthorHeaders{Diyi Liu, Yangsong Gu, and Lee Han}
\title{Estimating journey time for two-point vehicle re-identification survey with limited observable scope using 2-dimensional truncated distributions}

\author{%
  \textbf{Diyi Liu, Ph.D., Corresponding Author}\\
  Affiliations: Department of Civil and Environment Engineering\\
  University of Tennessee, Knoxville, Tennessee, USA\\
  Email: dliu27@vols.utk.edu\\
  \hfill\break
  \textbf{Yangsong Gu, Ph.D.}\\
  Affiliations: Department of Civil and Environment Engineering\\
  University of Tennessee, Knoxville, Tennessee, USA\\
  Email: ygu17@vols.utk.edu\\
  \hfill\break
  \textbf{Lee D. Han, Ph.D.}\\
  Affiliations: Department of Civil and Environment Engineering\\
  University of Tennessee, Knoxville, Tennessee, USA\\
  Email: lhan@utk.edu
}

\WordsPerTable{250}

\TotalWords{6272}

\begin{document}
\nolinenumbers
\maketitle




\section{Abstract}
In transportation, Weigh-in motion (WIM) stations, Electronic Toll Collection (ETC) systems, Closed-circuit Television (CCTV) are widely deployed to collect data at different locations. Vehicle re-identification, by matching the same vehicle at different locations, is helpful in understanding the long-distance journey patterns. In this paper, the potential hazards of ignoring the survivorship bias effects are firstly identified and analyzed using a truncated distribution over a 2-dimensional time-time domain. Given journey time modeled as Exponential or Weibull distribution, Maximum Likelihood Estimation (MLE), Fisher Information (F.I.) and Bootstrap methods are formulated to estimate the parameter of interest and their confidence intervals. Besides formulating journey time distributions, an automated framework querying the observable time-time scope are proposed. For complex distributions (e.g, three parameter Weibull), distributions are modeled in PyTorch to automatically find first and second derivatives and estimated results. Three experiments are designed to demonstrate the effectiveness of the proposed method. In conclusion, the paper describes a very unique aspects in understanding and analyzing traffic status. Although the survivorship bias effects are not recognized and long-ignored, by accurately describing travel time over time-time domain, the proposed approach have potentials in travel time reliability analysis, understanding logistics systems, modeling/predicting product lifespans, etc.

\hfill\break%
\noindent\textit{Keywords}:  vehicle re-identification, journey time distribution, travel time distribution, travel time reliability, travel time variability, applied truncated distribution, survivorship bias effects
\newpage

\section{1. Introduction \& Motivation}
The vehicle re-identification problem, defined as the problem of re-identifying the same vehicle at different observation sites, is claimed to be solved by license plate surveillance videos, license plate recognition (LPR) algorithms, and a clever, self-learning algorithm proposed by \citet{oliveira2012online, oliveira2013online} named automatic-LPR (ALPR), which easily achieves a precision over 95 \% for vehicle re-identification by matching license plate with minimized Levenshtein edit distances. This high performance was achieved based on the traditional image recognition techniques without the even better-performed deep learning algorithms involved \cite{polishetty2016next, hsu2012application}.

By applying the ALPR algorithm or other re-identification algorithms, vehicular travel/ journey time information along road segments between stations can be easily observed. The simplest form of such data collecting practice, referred to as the two-point survey, is to re-identify vehicles at two fixed located stations by matching license plates.

For two-point survey, two topics are of interests: (1) Evaluate the traffic congestion status over time (by monitoring travel time of re-identified vehicles between two stations) and (2) understand trucks’ travel behaviors and logistics operations underneath. For trucks, journey time can range from hours to days between two observation sites not far apart (e.g., 20 miles) due to logistics operations. Meanwhile, data collection time could be limited by legislation, labor costs, daylight levels, limited battery or disk space, etc. Thus, there exists a "survivorship bias" effects: given limited observation time windows at upstream/downstrream stations, vehicles with shorter journey times plus arrive early at upstream stations have a higher chance of being re-identified for those with a longer journey time and late arrival time.

\begin{figure}[hbt!]
\includegraphics[width=1.0\textwidth]{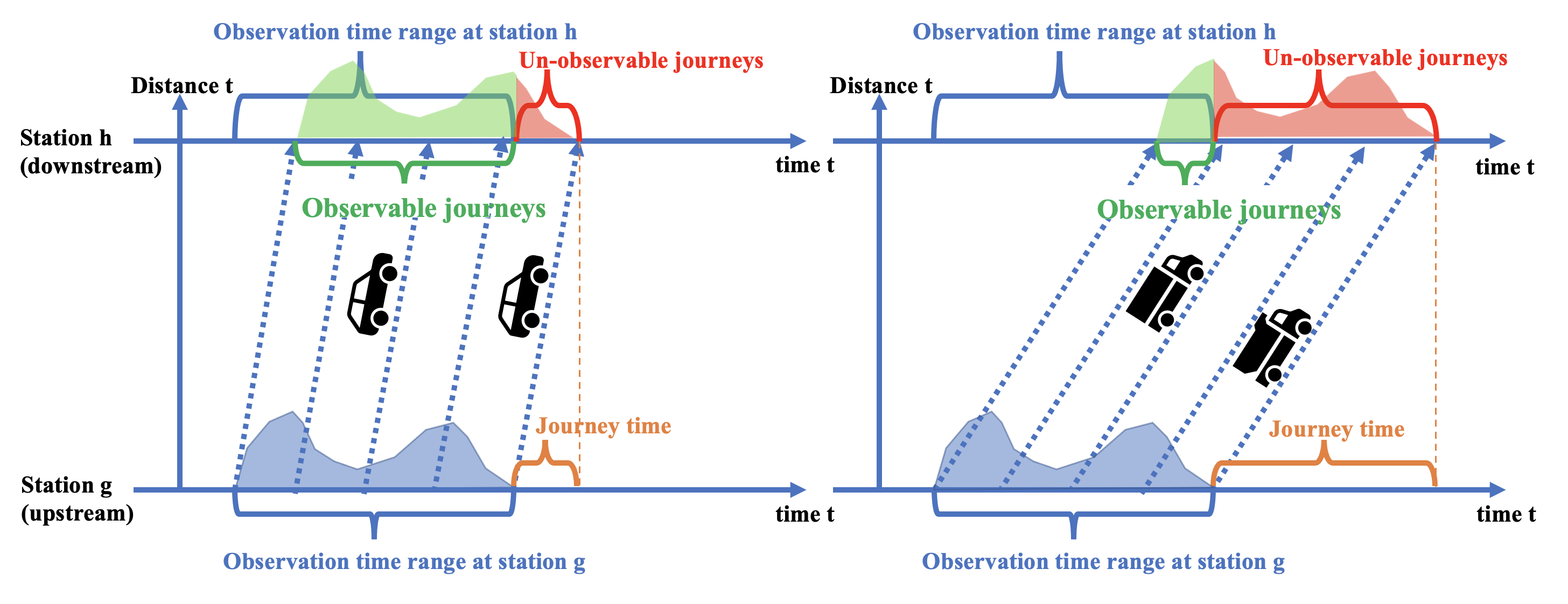}
\centering
\caption{Limited observable scope and the survivorship bias effects in two-point survey demonstrated on the time-space diagram}
\label{fig:0}
\end{figure}

\textbf{Figure \ref{fig:0}} shows a typical case of the survivorship bias effects in two-point survey where x-axis is the time range and y-axis corresponds to the distance along the travel direction. The histograms along station $g$'s and $h$'s x-axis represent the traffic counts arriving at station $g$ and $h$ over time of the day, respectively. The left diagram shows a group of small vehicles traveling at high speed, whereas the right diagram shows another group of trucks traveling at low speed. Given limited observation time range, there are un-observable vehicles for both groups (see red histograms at station $h$). Compared to small vehicles, a higher proportion of trucks arrives after the data collection ends (see red histograms) at downstream station $h$ since trucks travel slower than small vehicles. Ignoring such survivalship bias effects may severely underestimate truck's traffic volume.

\begin{figure}[hbt!]
\includegraphics[width=1.0\textwidth]{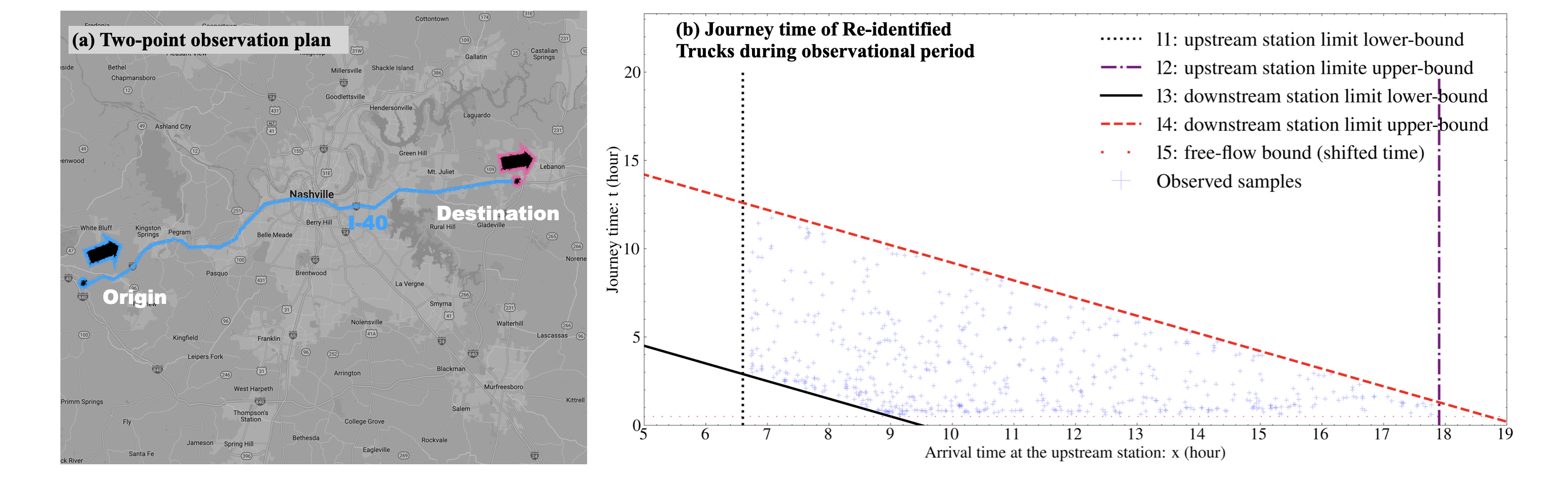}
\centering
\caption{An example of two-point survey re-identifying trucks}
\label{fig:1}
\end{figure}

Note that the two scenarios encapsulated in \textbf{Figure \ref{fig:0}} are largely simplified where each group of vehicles are assumed to have the same traveling speed (i.e., see the parallel dashed trajectory lines in time-space diagram). In practice, the average speed would follow some distributions and trucks may also reroute to other roads to finish their planned logistics schedules. To describe vehicle's stochastic behaviors, the time-time diagram comes into play. \textbf{Figure \ref{fig:1}} provides an example of a typical two-point survey with limited observation scope. \textbf{Figure \ref{fig:1}} (a) shows the eastbound I-40 highway corridor near Nashville, Tennessee along which a two-point survey for trucks are executed \cite{han2019automated}. Trucks' plates are recorded at both the origin (blue arrow) and destination (red arrow) station. Then, license plates from two stations are matched to re-identify trucks. \textbf{Figure \ref{fig:1}} (b) applied a time-time plot to demonstrate the re-identified vehicles using scatter plots. For \textbf{Figure \ref{fig:1}} (b), let x-axis denote the arrival time at the upstream station (origin) and y-axis denote the journey time in hours between two stations for each re-identified truck. Each blue cross in \textbf{Figure \ref{fig:1}} (b) represents a re-identified truck. By observation, it seems that all scatter-points are bounded by some lines, and there should be more unidentified data points outside of these lines, which will be discussed in the next paragraph.


In this paper, the observable zone refers to the range that can be observed from the time-time diagram. More information can be derived from \textbf{Figure \ref{fig:1}} (b), the time-time diagram. Five different straight lines (denoted from $l1$ to $l5$) jointly recognize the boundaries of the observable zone. Two vertical dashed lines (the densely dotted black line $l1$ and the dash-dotted purple line $l2$) identify the observation time ranges from 6:45 AM to 8:00 PM at the upstream station. By observation, not all trucks arrive between 6:45 AM to 8:00 PM at the upstream station can be re-identified. Only trucks between $l3$ and $l4$ (two parallel lines with slope of -1) can be identified. Besides, $l5$ (thin dotted red line) refers to the lower horizontal line corresponding the free flow travel time no greater than any truck's travel time.

\begin{figure}[!ht]
\includegraphics[width=\textwidth]{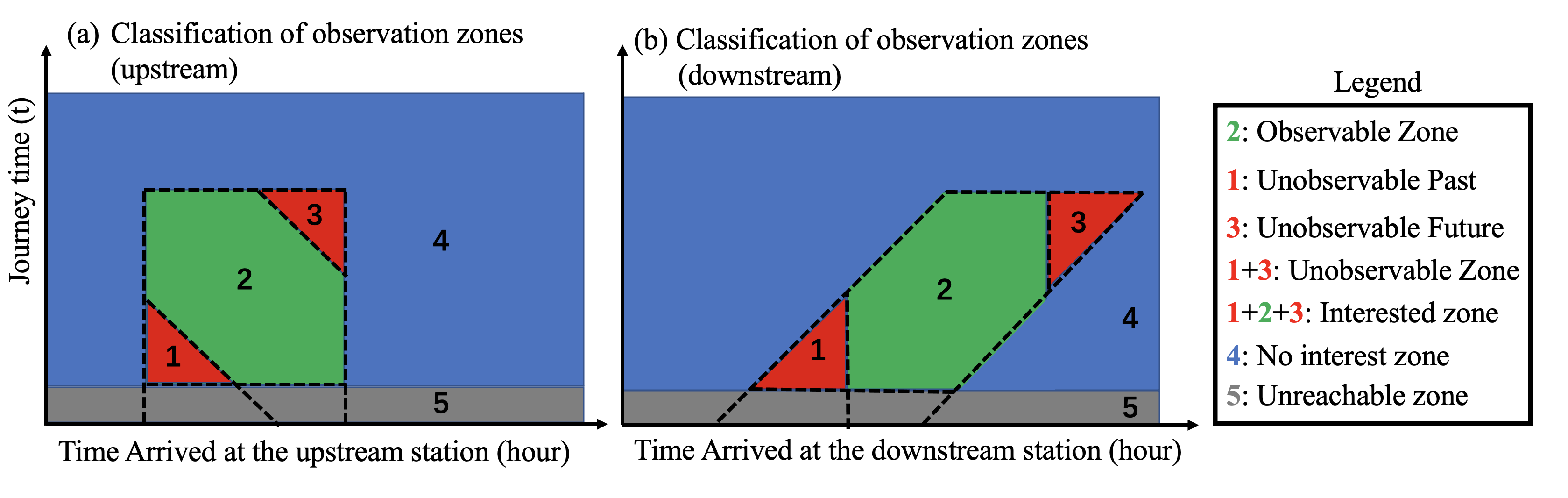}
\centering
\caption{Zone classification over time-time diagram (upstream-centric view)}
\label{fig:zone_classification}
\end{figure}

Representing data over the time-time domain, the case of the two-point survey problem is further generalized. \textbf{Figure \ref{fig:zone_classification}} (a) depicts a generalized case where the observation range is limited by the starting/ending times of the upstream/downstream observation stations. The x-axis corresponds to the arrival time of the day at the upstream station, and the y-axis corresponds to the journey time between both stations. The whole time-time diagram can then be classified into five different zones. The observer is only interested in studying the vehicles behavior within the "rectangular box" (i.e., Zone 1, Zone 2, and Zone 3). Zone 1 and Zone 3 are un-observable but are zones of interest. Zone 2 in green corresponds to the observable zone. Zone 4 is the "not interested" zone either because trucks arrive out of the upstream station’s observational range, or the journey time is too long to be modeled. No vehicles should fall into Zone 5 (in grey) because they cannot out-speed the free-flow speed.

Another useful time-time diagram is to plot the journey time vs. arrival time at the downstream station, as shown in \textbf{Figure \ref{fig:zone_classification}} (b). Zone 1 to Zone 5 also corresponds to the same Zone 1 to Zone 5 in \textbf{Figure \ref{fig:zone_classification}} (a). By observation, from \textbf{Figure \ref{fig:zone_classification}} (b), one can clearly see the reason why Zone 1 and Zone 3 are excluded from the observable zone: they fall out of the observation time range of the downstream station.

\begin{figure}[!ht]
\includegraphics[width=\textwidth]{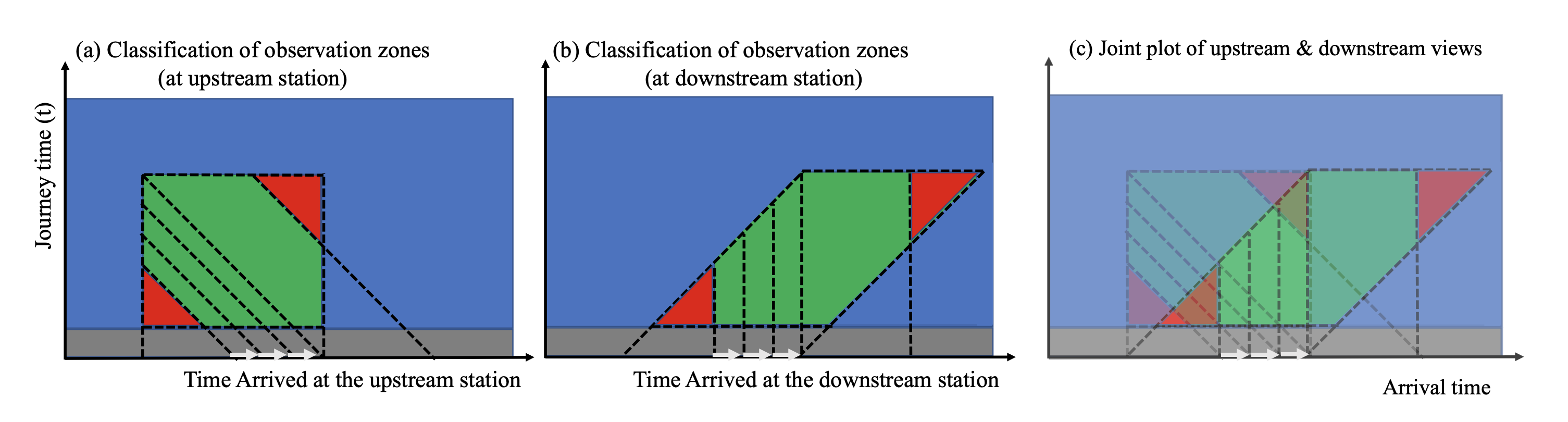}
\centering
\caption{The affine transformation relationship between upstream and downstream view (upstream-centric view)}
\label{fig:zone_classification2}
\end{figure}

To further understand the relationship between two views of journey time, the observable zones are split into several wedges by some parallel dash lines in \textbf{Figure \ref{fig:zone_classification2}}. By observation, one can use an affine transformation (i.e. 45 degrees clockwise transformation) to reach from \textbf{Figure \ref{fig:zone_classification2}} (a) to (b), as demonstrated in \textbf{Figure \ref{fig:zone_classification2}} (c) where (a)-(b) plots are overlaid and combined into one plot. The point is, since the affine transformation is one-to-one mapping function, any closed range in \textbf{Figure \ref{fig:zone_classification2}} (b) can be mapped to another closed range in \textbf{Figure \ref{fig:zone_classification2}} (a). Thus, modeling distribution densities on one sub-figure is enough to describe distribution densities on the other sub-figure with the help of affine transformation. For example, given a fitted probability density model on \textbf{Figure \ref{fig:zone_classification2}} (a), the number of re-identified vehicles at the downstream station can also be estimated by integrating the distribution over the slides piece by piece by integrating along those white arrows in \textbf{Figure \ref{fig:zone_classification2}} (a)-(c). One can also think of the series of white arrows as the downstream observer gradually postponing the downstream station's starting time. Thus, the theoretical arrival rate at the downstream station can be estimated just by modeling on the upstream station's time-time diagram.

\begin{figure}[!ht]
\includegraphics[width=\textwidth]{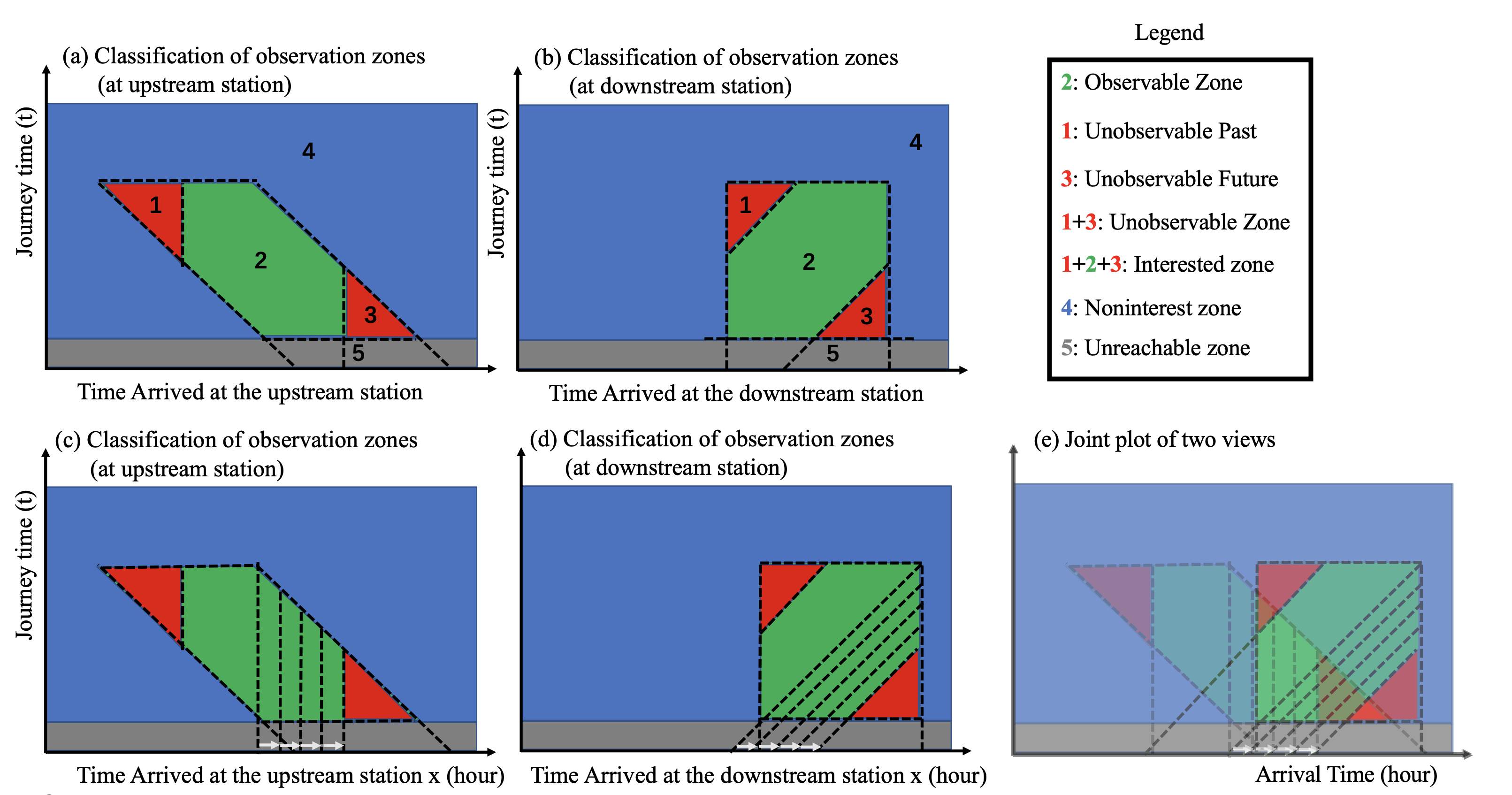}
\centering
\caption{The affine transformation relationship between upstream and downstream view (downstream-centric view)}
\label{fig:zone_classification3}
\end{figure}

Previously, the interested population are within a rectangular box for the upstream traffic, which is known as upstream centric view. Similarly, if one is interested in downstream arrivel time distribution, another set of observational zones are visualized as shown in \textbf{Figure \ref{fig:zone_classification3}}.

\begin{figure}[!ht]
\includegraphics[width=\textwidth]{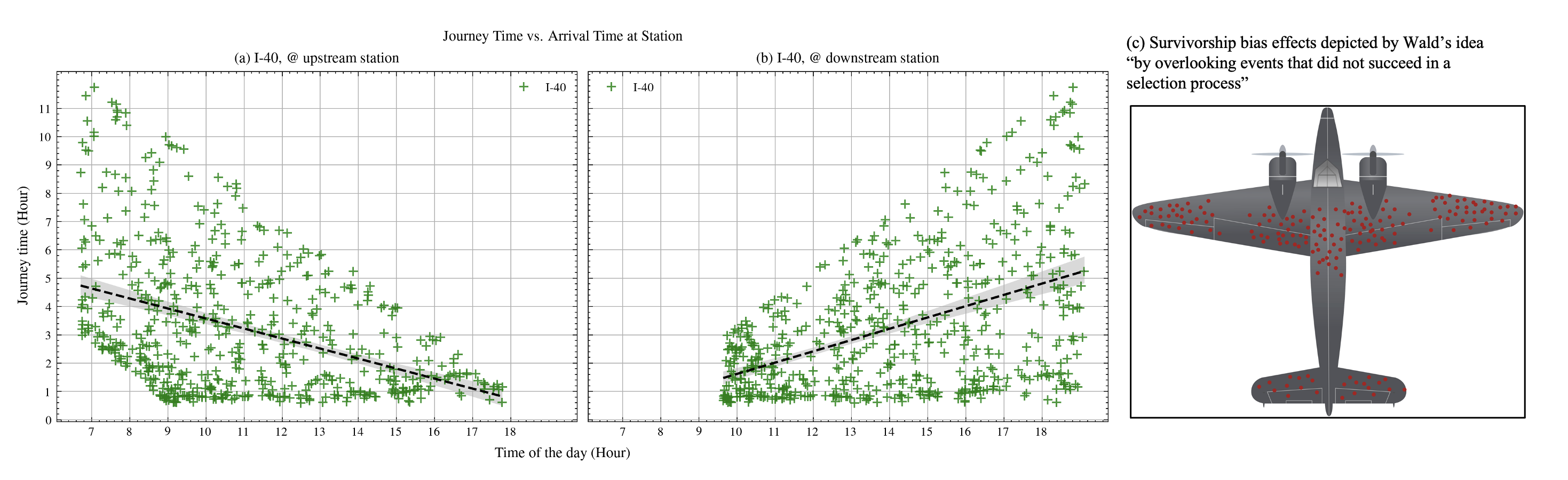}
\centering
\caption{A visualization of the survivorship bias effects}
\label{fig:ssb}
\end{figure}

Before deriving into math details, the hazards of ignoring the survivorship bias effect must be emphasized. One famous example of survivorship bias is the World War II airplane vulnerability estimation \cite{wald1980method} where planes are more likely to survive (thus observed) if vulnerable areas are not attacked, as shown in \textbf{Figure \ref{fig:ssb}} (c). For two-point survey, in \textbf{Figure \ref{fig:ssb}} (a)-(b), regression lines wrongly suggest that the journey time gets shorter and shorter at upstream locations, whereas the opposite case occurred at the downstream stations, all lead by ignoring the survivor-bias effect.

The remainder of this paper is structured as follows: In Section 2, the literature review section discusses the use of truncated distribution, two-dimensional distribution in transportation study. Also, the relationship between two-point observations, lifespan modeling, and survival analysis are briefly discussed. In Section 3, the two-point re-identification problem is formulated using truncated distribution over the 2-dimensional time-time space. In Section 4, the potential usages of proposed method in transportation studies are demonstrated using three experiments, that is: truck re-identification, route traffic status monitoring, and product-life modeling. In the end, the future potentials of using the 2-dimensional truncated distribution as well as identifying the survivorship bias effect are further discussed.

\section{2. Literature Review}
\subsection{2.1 Truncated distribution in transportation study}
One major application of truncated distributions in transportation study is to estimate travel speed. Truncated Normal/Log-normal distributions are usually assumed to better describe the distribution of travel time or speed. For example, \citet{cao2014modeling} used truncated distributions to model the traveling speed of in-motion vehicles in a signalized road section. \citet{wang2012speed} applied truncated Normal/Log-normal distribution to model travel speed/time. In most cases, most transportation studies focus on using truncated one-dimensional distribution modeling speed. In general, the probability density function (PDF) of such truncated distributions is:

\begin{linenomath}
\begin{equation}\label{eq:lr1}
f(t; \theta) = \frac{f_u(t; \theta)}{\int^{x_1}_{x_0} f_u(t; \theta) dx}, \;\; x_0 \leq{t} \leq{x_1}
\end{equation}
\end{linenomath}

where $t$, $\theta$ corresponds to the journey time and the model's paramters, respectively. $f_u(t;\theta)$ represents the PDF without truncation. The lower/upper bounds $x_0$ and $x_1$ are a fixed number, which can be estimated or known in advance given some assumptions.

\subsection{2.2 Time-time diagram in journey time monitoring}
In transportation, one related problem using time-time diagram is transit travel time reliability (TTTR). Similar to the two-point survey in \textbf{Figure \ref{fig:1}}, when analyzing travel time variability along a travel corridor, the departure time at the upstream station and the travel time can be plotted for each trip. Among all observations over time, the variations in travel time within a moving average window can be visualized by drawing some specific percentiles (e.g., 5th, 10th, 50th, 90th, 95th) over time, where a comparison between the travel time of the 90th percentile and the 50th percentile (i.e., the median) travel time can show the stability of transit service level \cite{kieu2015public, mazloumi2010using}. In practice, if the ratio becomes large (e.g., 2), then the transit system is considered to be unreliable. For different periods of comparison (e.g., day-to-day, month-to-month, during-peak/off-peak), different thresholds are used to evaluate the stability of the system.

\subsection{2.3 Parametric models in modeling duration of a status}
Compared to the Exponential distribution with constant failure rates (i.e., the memory-less property), the Weibull distribution, generalized from the Exponential distribution, can model time-varying failure rates. Thus, the Weibull distribution is frequently used in product reliability and lifespan modeling, mechanics monitoring, etc. \cite{surucu2009monitoring,zhang2011upper, mudholkar1996generalization, pham2007recent, carrasco2008generalized}. Besides the classic two-parameter and three-parameter Weibull distribution, more generalized versions of Weibull distribution are proposed over the last three decades to model time varying failure rates, especially the bathtub shaped failure rates \cite{mudholkar1996generalization, pham2007recent, carrasco2008generalized}. Several studies also demonstrated the versatility of Weibull distribution in reliability modeling \cite{zhang2011upper, kantar2015analysis}.

Several methods are all useful in estimating Weibull distributions. The MLE is the most frequently used one among them \cite{surucu2009monitoring,zhang2011upper, mudholkar1996generalization, pham2007recent, carrasco2008generalized}. Besides, Method of moments (MoM), graphical methods \cite{zhang2011upper}, and Bayesian estimation \cite{kaminskiy2005simple} are often adopted. Based on MLE, Fisher Information plays a critical role in determining the confidence intervals of parameters \cite{mudholkar1996generalization} for Weibull distributions.

\subsection{2.4 Different data censoring type under the contexts of survival analysis trials}
There seems to be some connections between two-point survey data and the censored data in survival analysis. For two-point survey, three time variables together compose the observed data: arrival time at the upstream station, arrival time at the downstream station, and the travel/journey duration between two stations. Given that, this observational plan may fit with the experimental concept of doubly censored data \cite{sun2006statistical}. One example of doubly censored data in medicine is to study the time difference between being infected by a virus (e.g., HIV) and showing symptoms (e.g., AIDS). However, unlike the AIDS example, many vehicles arrived at the upstream station will not travel along the corridor and reach the downstream station for the vehicle re-identification problem. To complicate things, the censoring time is also moving with vehicle's arrival time of the day.

\subsection{2.5 Summary}
Although no well-established methods can model the two-point survey problem, many ideas can be borrowed from previous studies. The time-time diagram can be used directly to describe the two-point survey data. Truncated Weibull/Exponential distribution can also be employed to model the journey time. Considering the time-dependent observing window, a good approach is to model journey time over the two-dimensional time-time space using truncated model.

\section{3. Methodology}  

\subsection{3.1 Modeling journey time distribution over time-time space}
Let $t$ and $x$ denote the journey time and the arrival time at the upstream stations respectively. Given the PDF of (un-truncated) distribution in 2-dimensional time-time space as $f_u(t, x; \theta)$, any two-dimensional truncated distribution bounded by closed region can be summarized in \textbf{Equation \ref{eq:core-1}}.

\begin{linenomath}
\begin{equation}\label{eq:core-1}
f(t, x; \theta) = \frac{f_u(t, x; \theta)}{\iint_S f_u(t, x; \theta) dS}
\end{equation}
\end{linenomath}
where $\theta$ denotes the set of parameters to be estimated, and $S$ represents  the support (i.e., observable zone) of the truncated distribution on the time-time diagram. Assuming the journey time $t$ and the arrival time $x$ are independent variables, the function can be decomposed as shown in \textbf{Equation \ref{eq:core-2}}:

\begin{linenomath}
\begin{equation}\label{eq:core-2}
f(t, x; \theta) \,\, \iidequal \,\, \frac{h(x) \cdot g(t; \theta)}{\iint_S h(x) \cdot g(t; \theta) dS} =
    \frac{h(x) \cdot g(t; \theta)}{\int_x \int^{b_u(x)}_{b_l(x)} h(x) \cdot g(t; \theta) dt dx} =
    \frac{h(x) \cdot g(t; \theta)}{\int_x  h(x) \int^{b_u(x)}_{b_l(x)}  g(t; \theta) dt dx}
\end{equation}
\end{linenomath}

Given vehicles' arrival time at $x$, the upper/lower observable journey time are given by $b_u(x)$ and $b_l(x)$, respectively. In \textbf{Equation \ref{eq:core-2}}, $h(\cdot)$ and $g(\cdot)$ denote the marginal distribution over the time of the day and over the journey time, respectively. Although the true $h(\cdot)$ is not known, it can be approximated by $w(\cdot)$, the observed arrival rate at the upstream station. Assuming $h(\cdot) \approx w(\cdot)$, the distribution becomes:

\begin{linenomath}
\begin{equation}\label{eq:core-25}
f(t, x; \theta) \approx \frac{w(x) \cdot g(t; \theta)}{\int_x  w(x) \int^{b_u(x)}_{b_l(x)}  g(t; \theta) dt dx}
\end{equation}
\end{linenomath}

Furthermore, one can assume $h(\cdot)$ stays about the same over time, the distribution $h(\cdot)$ can be simplified to the Uniform distribution. Thus, the function can be further simplified to:

\begin{linenomath}
\begin{equation}\label{eq:core-3}
f(t, x; \theta) 
    \approx \frac{c \cdot g(t; \theta)}{\int_x  c \cdot \int_{t(x)}  g(t; \theta) dt dx} 
    = \frac{g(t; \theta)}{\int_x \int^{b_u(x)}_{b_l(x)}  g(t; \theta) dt dx} 
    = \frac{g(t; \theta)}{\iint_S  g(t; \theta) dS}
\end{equation}
\end{linenomath}
where $c$ is a constant and the problem is philosophically reduced to a one-dimensional distribution but with the data dependent truncation bound. Thus, all parametric model for the two-point survey problem can be formulated using \textbf{Equation \ref{eq:core-1}-\ref{eq:core-3}}.

Although $g(t; \theta)$ can have many different formulations (e.g., Weibull, Exponential, etc.), only the standard 1-parameter Exponential distribution and the standard 2-parameter Weibull distribution are used in this study. Combined with two possible forms for $h(x)$, there are four different parametric models, as shown in \textbf{Table \ref{tab:four_scenarios}}.

\begin{table}[h]
\caption{Four Different Schemes Modeling the 2-dimensional Journey Time Problem}
\label{tab:four_scenarios}
    \centering
    \begin{tabular}{r | c | c}
        \hline
            \diagbox{Arrival time (horizontal) \\ distribution}{Journey time (vertical) \\ distribution} & Exponential & Weibull \\
        \hline
            Uniform distribution & Model 1 & Model 2 \\
        \hline
            Empirical distribution $w(t)$ & Model 3 & Model 4 \\
        \hline
    \end{tabular}
\end{table}

In the next section, the methods of estimating parameter of interests and their confidence intervals are introduced. First derivatives and second derivatives are necessary to estimate these parameters and their confidence intervals. As the form becomes more and more complicated, explicit solution becomes tedious or even non-exist. Thus, the authors take advantage of autograding softwares (i.e., PyTorch) to numerically compute the first and secondary derivatives automatically to find parameter of interests.

\subsection{3.2 Using MLE \& Fisher Information to estimate parameters and their confidence intervals}
For Model 1 in \textbf{Table \ref{tab:four_scenarios}}, under the assumption that journey times and arrival time follows an Exponential distribution and a Uniform distribution respectively, the PDF of the truncated distribution becomes:

\begin{linenomath}
\begin{equation}
f(t, x, \lambda) = \frac{\lambda e^{-\lambda t}}{\int_{x_s}^{x_e} \int_{b_l}^{b_u} \lambda e^{-\lambda t} dt dx} = \frac{\lambda e^{-\lambda t}}{\int_{x_s}^{x_e} e^{-\lambda b_l} - e^{-\lambda b_u} dx} \;.
\end{equation}
\end{linenomath}

where $x_s$, $x_e$ represent the starting/ending recording time at the upstream station. Then, the joint likelihood function for all $n$ observations can be expressed as:

\begin{linenomath}
\begin{equation} \label{Eq:l}
\mathcal{L}(\lambda; \vec{t}, \vec{x}) = \prod_{i=1}^{n} f(t_i, x_i; \lambda) = \frac{\lambda^n e^{-\lambda \sum_i{t_i}}}{ \left[ \int_{x_s}^{x_e} e^{-\lambda b_l} - e^{-\lambda b_u} dx \right]^{n} }
\end{equation}
\end{linenomath}

\noindent where the value of $b_l$, $b_u$ for each observation solely depends on the vehicle's passage time $x$ at upstream station. Also, observed data set recorded as $\{(t_1,x_1), \ldots, (t_n, x_n)\}$ , and ${t_0, t_1, \ldots, t_n}$ and ${x_0, x_1, \ldots, x_n}$ are represented as $\vec{t}$ and $\vec{x}$, respectively. By taking logarithm, the log-likelihood function (LL) is derived as follows:

\begin{linenomath}
\begin{flalign}
\log \mathcal{L}(\lambda; \vec{t}, \vec{x}) = n \log{\lambda} - \lambda \sum_{i=1}^{n} t_i - n \log{ \int_{x_0}^{x_1} (e^{-\lambda b_l} - e^{-\lambda b_u}) dx } \; .
\end{flalign}
\end{linenomath}

\noindent
Taking the first partial derivative with respect to $\lambda$ over LL gives:

\begin{linenomath}
\begin{equation} \label{eq:8}
\frac{\partial \log \mathcal{L}(\lambda; \vec{t}, \vec{x})}{\partial \lambda}
 = \frac{n}{\lambda} - \sum_{i=1}^{n}{t_i} - n \cdot \frac{\int_{x_0}^{x_1} b_u \cdot e^{-\lambda b_u} - b_l \cdot e^{-\lambda b_l} dx}{ \int_{x_0}^{x_1} (e^{-\lambda b_l} - e^{-\lambda b_u} ) dx }
\end{equation}
\end{linenomath}

where the order of integration and differentiation is exchanged. For each observed data/ vehicle with upstream arrival time $x$, the lower and upper bounds of the truncated distribution (i.e., $b_l, b_u$) are known. Thus, one can numerically integrate out the integral parts in \textbf{Equation \ref{eq:8}}. By solving $\frac{\partial \log \mathcal{L}(\lambda; \vec{t}, \vec{x})}{\partial \lambda}=0$ using an existing equation solving algorithm (e.g., Newton's method), the root $\hat{\lambda}$ becomes a numerical Maximum Likelihood Estimator of $\lambda$.

\subsection{3.3 Estimate Confidence Intervals by considering Fisher Information}

After using MLEs to estimate parameters, Fisher Information can be applied to estimate confidence intervals. To simplify the expressions in \textbf{Equation \ref{eq:8}}, let: 

\begin{linenomath}
\begin{equation}\label{eq:x}
u = \int_{x_0}^{x_1} (e^{-\lambda b_l} b_l - e^{-\lambda b_u} b_u) dx \text{, }v = \int_{x_0}^{x_1} (e^{-\lambda b_l} - e^{-\lambda b_u} ) dx \;.
\end{equation}
\end{linenomath}

\noindent Then, the second derivative of LL can be expressed as: 
\begin{linenomath}
\begin{equation} \label{eq:10}
\frac{\partial^2 \log \mathcal{L}(\lambda; \vec{t},\vec{x})}{\partial \lambda^2}
 = \frac{-n}{\lambda^2} - n \cdot \frac{ \int_{x_0}^{x_1} (e^{-\lambda b_l} b_l^2 - e^{-\lambda b_u} b_u^2) dx \cdot v-u \cdot u}{v^2}
\end{equation}
\end{linenomath}

where the first part $\frac{-n}{\lambda^2}$ is just the Fisher Information for the Exponential distribution without truncation. The second part decrease the total carried information by each observation because distribution's support is limited. The Fisher Information for $n$ independent and identically distributed (i.i.d.) observations, by definition, can be expressed as:
\begin{linenomath}
\begin{equation} \label{Eq11.}
    I_n(\lambda) = -E_{\lambda}\left[\frac{\partial^2 \log \mathcal{L}(\lambda; \vec{t},\vec{x})}{\partial \lambda^2}\right]
    = -E\left[\frac{\partial^2 \log \mathcal{L}(\lambda; \vec{t}, \vec{x})}{\partial \lambda^2} \;\middle|\; \lambda\right] = n \cdot I(\lambda)
\end{equation}
\end{linenomath}

where $I(\lambda)$ denotes the Fisher Information given by one single observation. Given observed data set recorded as $\{(t_1,x_1), \ldots, (t_n, x_n)\}$, by definition, the total observed Fisher Information becomes:

\begin{linenomath}
\begin{equation} \label{Eq11.1}
    \widehat{I}_n(\hat{\lambda}) = 
    - \sum_{i=1}^{n} \left[ \frac{\partial^2 \log \boldsymbol\ell(\lambda; t_i, x_i)}{\partial \lambda^2} \right]
\end{equation}
\end{linenomath}

By Strong Law of Large Number (SLLN) \cite{casella2021statistical}, there holds:

\begin{linenomath}
\begin{equation} \label{Eq11.2}
    \frac{1}{n} \cdot \widehat{I}_n(\hat{\lambda}) \overset{p}{\longrightarrow} I(\lambda_0) \;\; \text{, as} \; n \rightarrow \infty \;\; .
\end{equation}
\end{linenomath}
where $\lambda_0$ denotes the true value of the parameter. Under some mild conditions, the maximum likelihood estimator $\lambda$ has the following property:

\begin{linenomath}
\begin{equation} \label{Eq11.6}
    \sqrt{ I_n(\lambda_0)} \cdot (\hat{\lambda} - \lambda_0) \overset{d}{\longrightarrow} N(0, 1)  \;\; \text{, as} \; n \rightarrow \infty \;\; .
\end{equation}
\end{linenomath}
Finally, by applying Slutsky's Theorem \cite{casella2021statistical} combined with \textbf{Equations \ref{Eq11.1}-\ref{Eq11.6}}:

\begin{linenomath}
\begin{equation} \label{Eq11.7}
    \sqrt{\widehat{I}_n(\hat{\lambda})} \cdot  (\hat{\lambda} - \lambda_0) \overset{d}{\longrightarrow} N(0, 1)  \;\; \text{, as} \; n \rightarrow \infty \;\; .
\end{equation}
\end{linenomath}

\noindent The estimated parameter $\hat{\lambda}$ asymptotically follows a Normal distribution as the data size $n$ increases. Consequently, the standard error follows \textbf{Equation \ref{Eq12}}:
\begin{linenomath}
\begin{equation} \label{Eq12}
    \text{s.e.($\hat \lambda$)} = \sqrt{\frac{1}{\widehat{I}_n(\hat \lambda)}} =   \widehat{I}_n^{-0.5}(\hat \lambda);
\end{equation}
\end{linenomath}

and the $(1-\alpha)$th confidence interval is given as:
\begin{linenomath}
\begin{equation}\label{Eq13}
    \left(\hat{\lambda} - z^{(1-\alpha/2)} \cdot \widehat{I}_n^{-0.5}(\hat \lambda), \;\; \hat{\lambda} + z^{(1-\alpha/2)} \cdot \widehat{I}_n^{-0.5}(\hat \lambda)\right) \;\; .
\end{equation}
\end{linenomath}

Besides, when multiple unknown parameters are involved, Fisher Information can be generalized to Fisher Information Matrix (see page 265-267 from \textit{The Elements of Statistical Learning} \cite{hastie2009elements}). For Weibull distribution, the derivation steps are given in sub-section 3.5 for the interested reader.

\subsection{3.4 Estimating Confidence Interval of Journey Time on top of MLE estimates using Bootstrap method}
Both the Fisher Information and Bootstrap methods can be applied to estimate confidence intervals based on MLE. However, Bootstrap is preferred because: (1) it requires no extra mathematical derivation; (2) it can easily be applied to complicated distributions where mathematical derivation is impossible.

\begin{figure}[!ht]
    \centering
    \includegraphics[width=\textwidth]{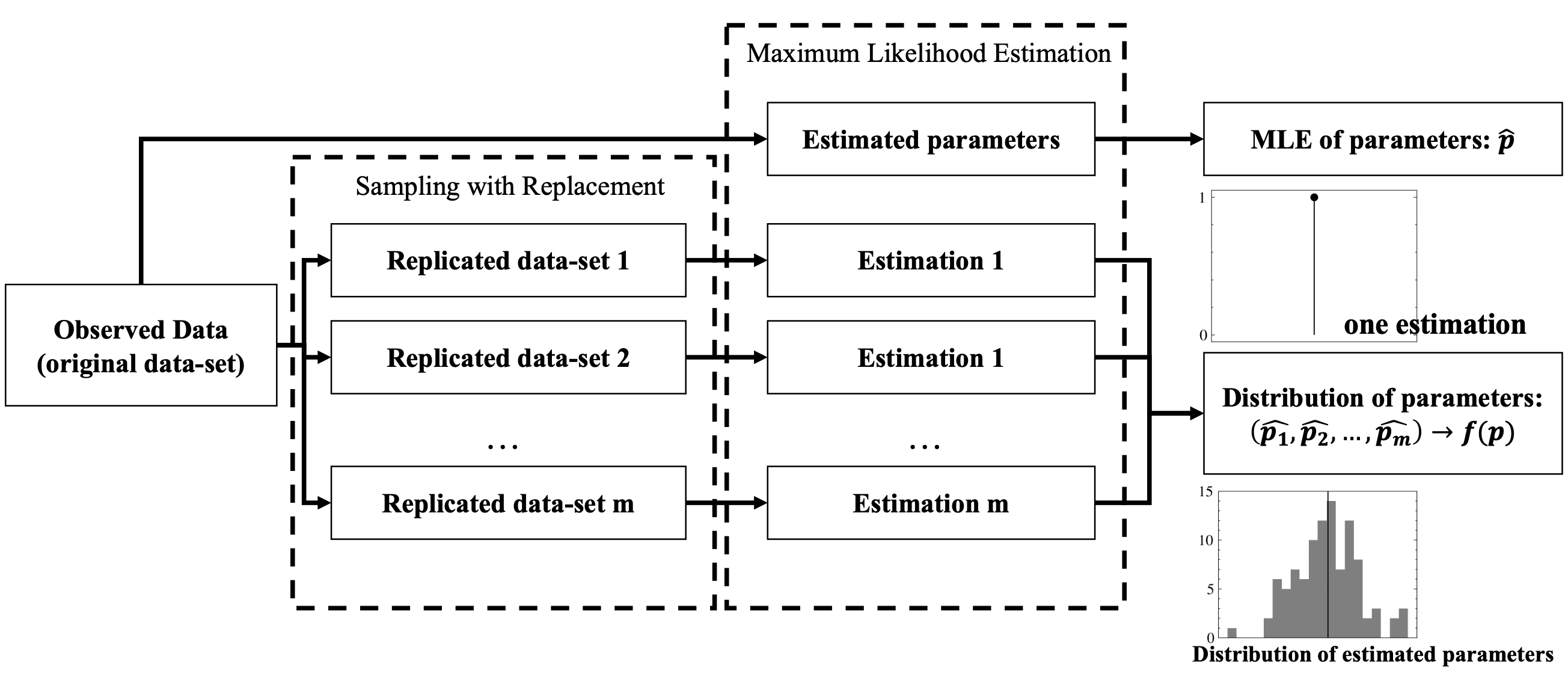}
    \caption{Procedures in estimating parameters using MLE and estimating the confidence interval on top of MLE using Bootstrap Method or Fisher Information}
    \label{fig:bootstrap}
\end{figure}

From the statistical view, the data sampling process can be imitated by redrawing samples with replacements from the observed data \cite{hesterberg2011bootstrap, hastie2009elements}. \textbf{Figure \ref{fig:bootstrap}} demonstrates the process by re-drawing samples with replacement for $k$ times ($k$ equals the size of all observed data) to form a replicated data-set. Each replicated data-set can provide an estimation for parameters and statistics of interest (e.g., the mean journey time). This process is then called $m$ times to give $m$ estimates of parameters, and their empirical histograms converge to the true distribution of the target parameters. \cite{hesterberg2011bootstrap, hastie2009elements}.

\subsection{3.5 Estimate 2-dimensional truncated distribution (given Weibull distributed journey time)}

The PDF of the two-parameter Weibull distribution is:
\begin{linenomath}
\begin{equation}
f(t; k, \lambda) = \begin{cases}
    \frac{k}{\lambda} (\frac{t}{\lambda})^{k-1}  e^{- (\frac{t}{\lambda})^k }, & \text{if } t \geq{0}  \\
    0, & \text{otherwise}
    \end{cases}
\end{equation}
\end{linenomath}
where the median and mean values are $\lambda \Gamma{(1 + 1/k)}$ and $\lambda (\ln 2)^{1/k}$, respectively. Under the assumption that journey time $t$ follows a Weibull distribution over time and arrival time at upstream station $x$ follows the Uniform distribution, the joint PDF becomes:

\begin{linenomath}
\begin{equation}
f(t, x; k, \lambda) = \frac{ \frac{k}{\lambda} (\frac{t}{\lambda})^{k-1} e^{- (\frac{t}{\lambda})^k } }{\int_{x_s}^{x_e} \int_{b_l}^{b_u} \, \frac{k}{\lambda} (\frac{t}{\lambda})^{k-1} e^{- (\frac{t}{\lambda})^k }\, dt dx} = \frac{ \frac{k}{\lambda} (\frac{t}{\lambda})^{k-1} e^{- (\frac{t}{\lambda})^k } }{ \int_{x_s}^{x_e} \, e^{-(\frac{b_l}{\lambda})^k} - e^{-(\frac{b_u}{\lambda})^k} \, dx } \, , \;\; \text{for } t \geq{0}
\end{equation}
\end{linenomath}
where $k$, $\lambda$ denotes the shape and the scale parameters of the distribution respectively. The LL function for one observation becomes:
\begin{linenomath}
\begin{equation}
    \begin{split}
        \Lambda & = \log{f(t, x; k, \lambda)} \\
        & = \log{(\frac{k}{\lambda}) + (k-1) \log{(\frac{t}{\lambda}) - (\frac{t}{\lambda})^k - \log{[\int_{x=x_s}^{x_e} e^{-(\frac{b_l}{\lambda})^k} - e^{-(\frac{b_u}{\lambda})^k} dx]} }} \\
        & = \log{k} - \log{\lambda} + (k-1) \log{t} - (k-1) \log{\lambda} - (\frac{t}{\lambda})^k - \log{[\int_{x=x_s}^{x_e} e^{-(\frac{b_l}{\lambda})^k} - e^{-(\frac{b_u}{\lambda})^k} dx]} \\
        & = \log{k} - k \log{\lambda} + (k-1) \log{t} - (\frac{t}{\lambda})^k - \log{[\int_{x=x_s}^{x_e} e^{-(\frac{b_l}{\lambda})^k} - e^{-(\frac{b_u}{\lambda})^k} dx]} \;\; .
    \end{split}
\end{equation}
\end{linenomath}

\noindent where the value of $b_l$, $b_u$ depends on the passage time $t$. To estimate parameteres using MLE, the first partial derivative function with respect to both $\lambda$ and $k$ are given in \textbf{Equation \ref{app:23}-\ref{app:24}}:

\begin{linenomath}
\begin{equation} \label{app:23}
    \begin{split}
        \Lambda_{\lambda} & = \frac{\partial \Lambda}{\partial \lambda} \\
        & = - \frac{k}{\lambda} + t^k \cdot k \cdot \lambda^{-(k+1)} - (k \cdot \lambda^{-k-1}) \cdot \frac{ \int_{x_e}^{x_s} \, e^{-(\frac{b_l}{\lambda})^k} \cdot b_l^k  - e^{-(\frac{b_u}{\lambda})^k} \cdot b_u^k \, dx }
        { \int_{x_s}^{x_e} \, e^{-(\frac{b_l}{\lambda})^k} - e^{-(\frac{b_u}{\lambda})^k} \, dx } \;\; \text{, \; and}
    \end{split}
\end{equation}
\end{linenomath}

\begin{linenomath}
\begin{equation} \label{app:24}
    \begin{split}
        \Lambda_k & = \frac{\partial \Lambda}{\partial k} \\
        & = \frac{1}{k} + \log{x} - \log{\lambda} - k (\frac{t}{\lambda})^{(k-1)} +
        \log{k} \cdot \frac{ \int_{x_0}^{x_1} \, e^{-(\frac{b_l}{\lambda})^k} \cdot  (\frac{b_l}{\lambda})^{k} - e^{-(\frac{b_u}{\lambda})^k} \cdot (\frac{b_u}{\lambda})^{k} \, dx }
        { \int_{x_0}^{x_1} \, e^{-(\frac{b_l}{\lambda})^k} - e^{-(\frac{b_u}{\lambda})^k} \, dx } \;\; .
    \end{split}
\end{equation}
\end{linenomath}

Note that the order of integration and differentiation is ex-changed above. The observed Fisher information is just the second derivatives of the LL function. For the 2-parameter Weibull distribution formulated above, the Fisher Information Matrix for one observation is:

\begin{equation}
    I(k,\lambda) = - \begin{bmatrix}
        \Lambda_{kk} & \Lambda_{k\lambda} \\
        \Lambda_{\lambda k} & \Lambda_{\lambda \lambda}
        \end{bmatrix} = - \begin{bmatrix}
        \frac{\partial^2 \Lambda}{\partial k^2} & \frac{\partial^2 \Lambda}{\partial k \partial \lambda} \\
        \frac{\partial^2 \Lambda}{\partial \lambda \partial k} & \frac{\partial^2 \Lambda}{\partial \lambda^2}
        \end{bmatrix} \;\; .
\end{equation}

For all observations, the Fisher Information is simply the summation of information of all data, as given in \textbf{Equation \ref{app:26}}:
\begin{equation} \label{app:26}
    I_n(k,\lambda) = \sum_{i=1}^{n}{I(k,\lambda)} = n \cdot I(k,\lambda) \;\; .
\end{equation}

Thus, for distribution with parameters $\vec{\theta} = (k, \lambda)$ and its MLE estimated value as $\hat{\theta} = (\hat k, \hat \lambda)$ , the distribution of parameters are asymptotically following a multivariate (2-dimensional) Normal distribution:
\begin{equation}
    \hat{\theta} \approx N_2(\hat{\theta}, \frac{1}{I_n(\hat{\theta})}) \;\; .
\end{equation}

Note that given the equations of the first derivative, the second derivative at $\hat{\theta}$ can also be estimated accordingly using numerical methods. For example:

\begin{equation}
    g'(x) = \lim_{\delta x \rightarrow{0}} \frac{g(x+\delta x) - g(x)}{\delta x} \approx \frac{g(x+\Delta x) - g(x)}{\Delta x}  \;\; , 
\end{equation}

where $g(x)$ is the function to take derivative and $\Delta x$ is a small value (e.g., $1\times 10^{-6}$). For more complicated distributions, the Bootstrap method is preferred in estimating confidence intervals.

\subsection{3.6 Evaluation methods}
Model evaluation is needed after fitting the parametric model. In this study, Kolmogorov–Smirnov (K-S) test is applied to compare results between observed CDF and emphrical CDF functions \cite{degroot2012probability}. The K-S test statistic is summarized as:

\begin{linenomath}
\begin{equation}
    D_n = \sup_x |F_n(x) - F(x)|
\end{equation}
\end{linenomath}

where $D_n$ is a statistic measuring the maximum distance between fitted CDF and empirical CDF given observed data. Under the null hypothesis that the fitted distribution can generate the observed data, a hypothesis test can be applied using p-value or critical value.


\section{4. Case Study \& Experiments}

\begin{table}[!ht]
\caption{Two Different Study Cases for Monte Carlo Simulation}
    \centering
    \begin{tabular}{|r|c|c|}
        \hline
        Study cases & Case 1 & Case 2 \\
        \hline
        Upstream observation period & \multicolumn{2}{c|}{6-9 AM} \\
        \hline
        Downstream observation period & \multicolumn{2}{c|}{7-10 AM} \\
        \hline
        Journey time distribution (hour) & Exponential$(\lambda=2)$ & Weibull($k=0.75, \lambda=2$) \\
        \hline
        Arrival distribution (hour) & \multicolumn{2}{c|}{Uniform} \\
        \hline
        Mean travel time (hour) & $\lambda=2$ & $ \lambda \cdot \Gamma{(1+\frac{1}{k})} = 2.38$ \\
        \hline
    \end{tabular}
\label{tab:exp_data}
\end{table}

\begin{figure}[!ht]
\centering
\includegraphics[width=\textwidth]{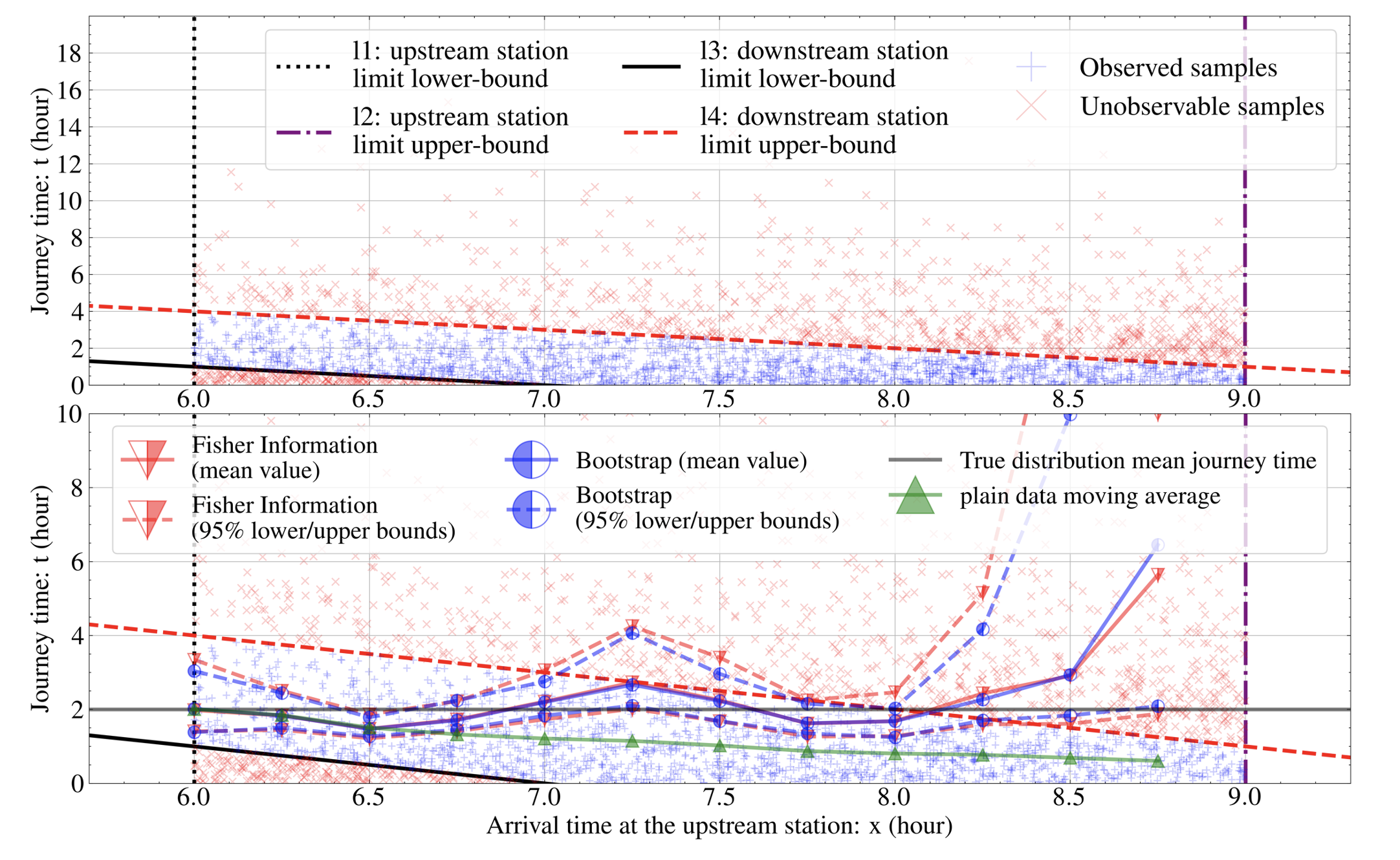}
\caption{An example of one Monte Carlo simulation run (Case 1)}
\label{fig:sim_case1}
\end{figure}

\subsection{4.1 Case Study: Analyzing truck behaviors over traffic corridor}
\begin{figure}[!ht]
\includegraphics[width=\textwidth]{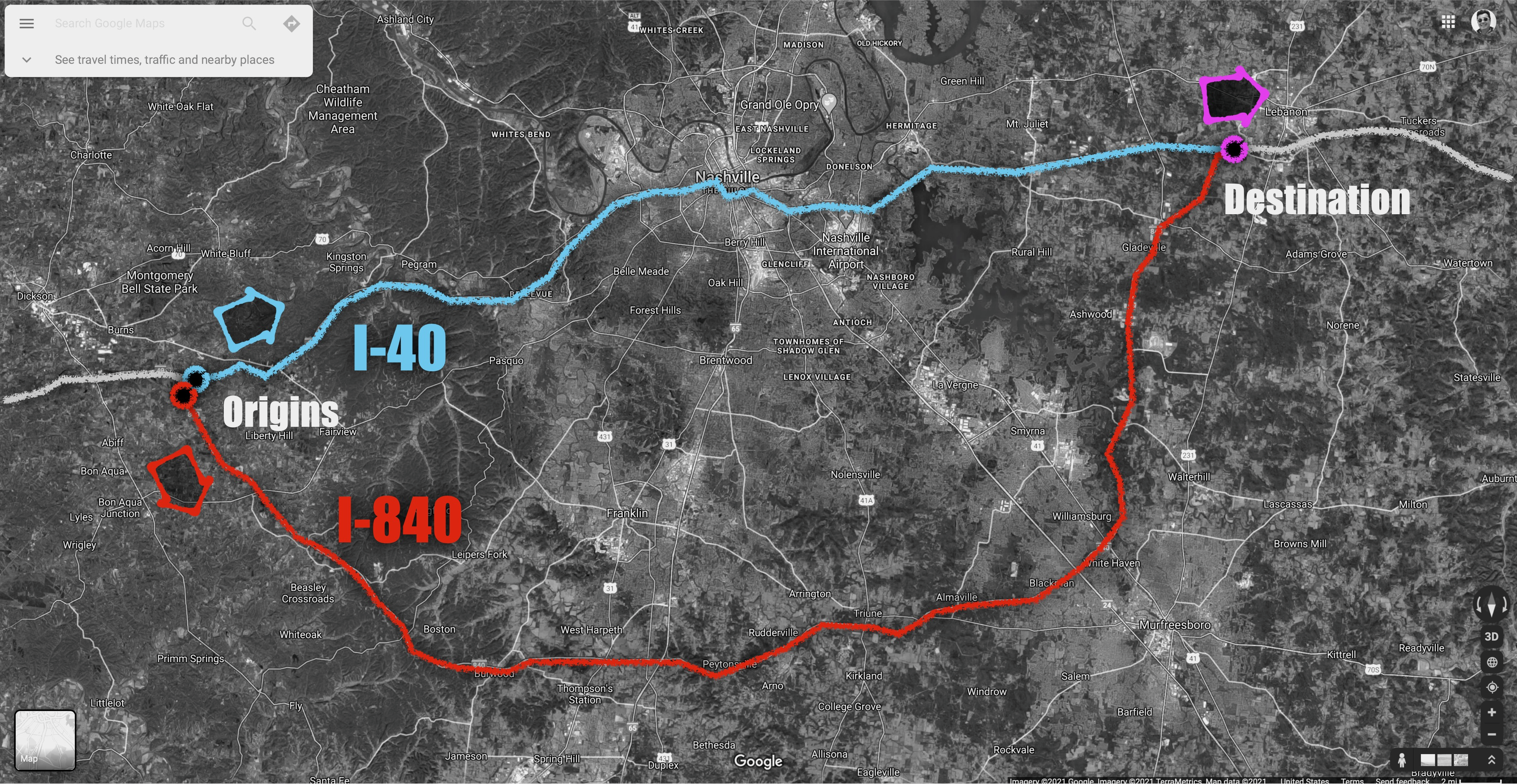}
\centering
\caption{Two possible eastbound travel routes near Nashville, Tennessee}
\label{fig:stations}
\end{figure}

After running Monte Carlo simulation to validate the proposed method, the same modeling approach is used to understand the truck’s journey time over a busy corridor in Nashville, Tennessee. \textbf{Figure \ref{fig:stations}} demonstrates the three established sites to identify trucks traveling eastbound along the corridor. Two separate two-point surveys are applied on I-40 and on I-840 during the same day, and the meta-information about this survey is summarized in \textbf{Table \ref{tab:meta}}  \cite{han2019automated}. The length of I-40 and I-840 are 60 miles and 78 miles, respectively. The starting/ending time of each station is also given in \textbf{Table \ref{tab:meta}}. While I-40 is shorter in distance, the speed limit of I-40 (55-70 mph) is slightly lower than that of I-840 (70 mph). Compared to the I-40 highway going through downtown Nashville, the I-840 corridor has the potential benefit of avoiding congestion during peak hours through a suburban region. In short, investigating truck’s journey time difference along two routes would be helpful in understanding truck behaviors.

\begin{table}[!ht]
\caption{Meta-information of the corridor case study}
    \centering
    \begin{tabular}{|c|c|c|}
        \hline
        Corridor name & Length (mile) & Speed limit (mph) \\
        \hline
        I-40 corridor & 60  & 55-70 \\
        \hline
        I-840 corridor & 78 & 70 \\
        \hline\hline
        Station name & Start time & End time \\
        \hline
        I-40 upstream station & 6:36AM & 5:54PM \\
        \hline
        I-840 upstream station & 7:12AM & 5:54PM \\
        \hline
        I-40/840 downstream station & 9:30AM & 7:12PM \\
        \hline
    \end{tabular}
\label{tab:meta}
\end{table}

\begin{figure}[!ht]
\includegraphics[width=\textwidth]{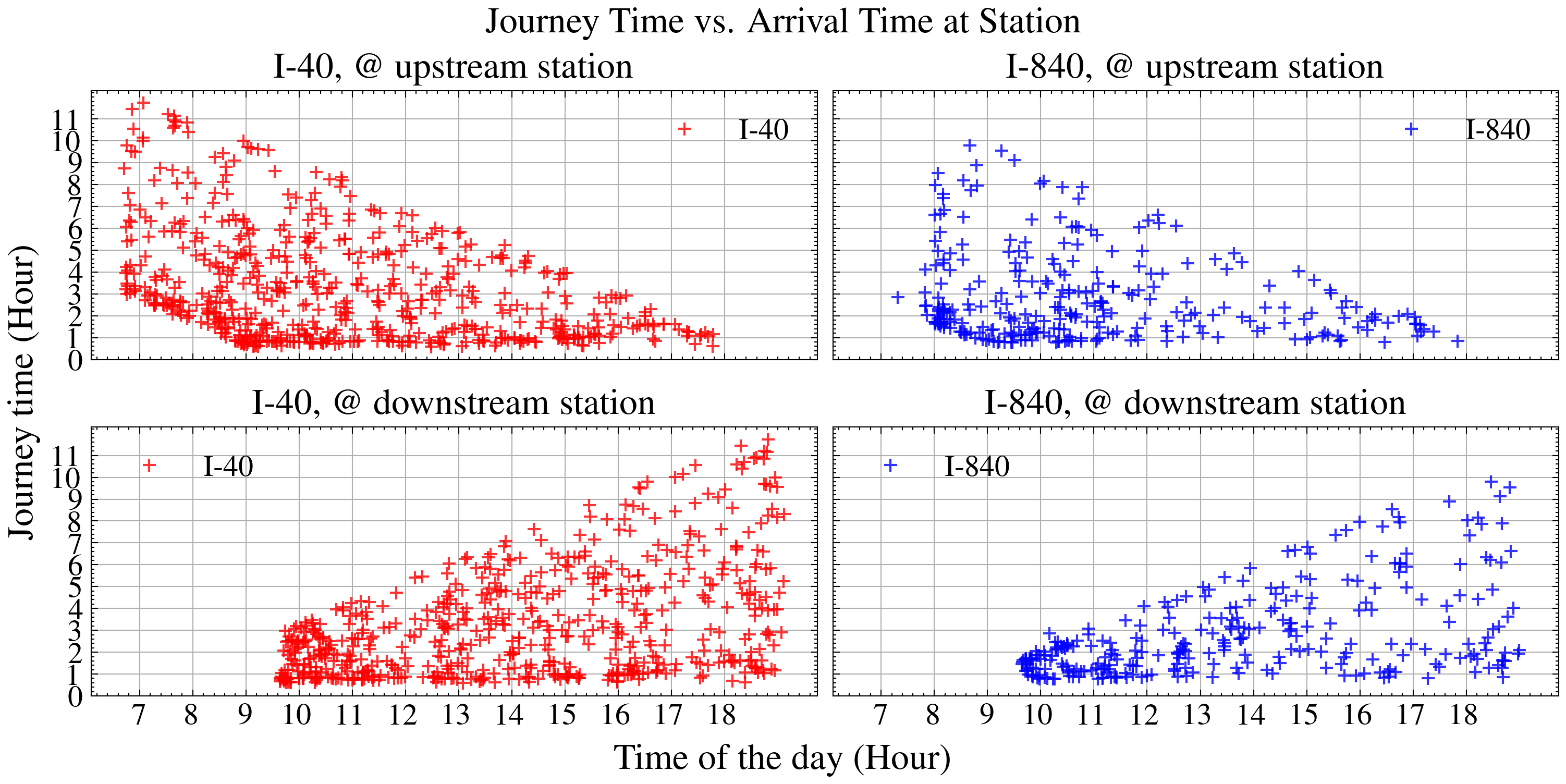}
\centering
\caption{Time-time plot of the I-40/I-840 corridor}
\label{fig:case_study_tt}
\end{figure}

\textbf{Figure \ref{fig:case_study_tt}} gives all observed data over the time-time plot (journey time vs. arrival time) for each traveling route. The top row of sub-figures provides the time-time plot where x-axis denotes the upstream arrival time. The bottom row of sub-figures provides time-time plot where x-axis denotes the downstream arrival time.


By modeling the journey time using proposed methods, one can get a better understanding of truck's behavior on each route. For each corridor, four models are fitted: (1) Exponential-Uniform Model; (2) Exponential-Empirical Model; (3) Weibull-Uniform Model; (4) Weibull-Empirical Model. The journey time between upstream and downstream stations (i.e., the "vertical" distribution) are modeled as Exponential/Weibull distributions. The arrival time at the upstream station (i.e., the "horizontal" distribution) is modeled as Uniform/Empirical distribution. The empirical distribution is estimated by interpolating the truck arrival rate over time at the upstream station. Although not all trucks identified at the upstream station will stick to the route and arrive at the downstream station, it is assumed that the proportion of trucks sticking to the corridor is around a fixed rate, and the Empirical model gives a good approximation.

\textbf{Table \ref{tab:est_results}} gives the estimated results for all models and all scenarios. All confidence intervals concentrate in a relatively small range, which is a good sign that the observed data have provided enough information to fit those models. While different models give different estimates of the mean journey time, the mean journey time is significantly smaller in I-840 compared to I-40 given the same modeling assumption.

\newpage

\begin{table}[!ht]
\caption{A Summary of Modeling Results}
\label{tab:est_results}
\scriptsize
    \begin{longtable}[]{@{}lllll@{}}
    \toprule
    & I-40 & & & \\
    \hline
    & Model 1 & Model 2 & Model 3 & Model 4 \\
    & Exponential + Uniform & Exponential + Frequency & Weibull + Uniform &
    Weibull + Frequency \\
    Mean Journey Time & 7.57 & 7.80 & 4.85 & 4.80 \\
    95\% C.I. (Fisher Information) & {[}6.08, 10.00{]} & {[}6.24, 10.38{]} &
    N/A & N/A \\
    95\% C.I. (Bootstrap) & {[}6.92, 10.49{]} & {[}6.69, 9.76{]} & {[}4.18,
    5.80{]} & {[}4.26, 5.56{]} \\
    \hline
    & I-840 & & & \\
    \hline
    Mean Journey Time & 3.39 & 2.84 & 3.45 & 3.15 \\
    95\% C.I. (Fisher Information) & {[}2.82, 4.23{]} & {[}2.44, 3.41{]} &
    N/A & N/A \\
    95\% C.I. (Bootstrap) & {[}3.02, 4.10{]} & {[}2.89, 4.36{]} & {[}3.03,
    3.91{]} & {[}2.86, 3.40{]} \\
    \bottomrule
    \addtocounter{table}{-1} 
    \end{longtable}
\end{table}

\begin{figure}[!th]
\includegraphics[width=\textwidth]{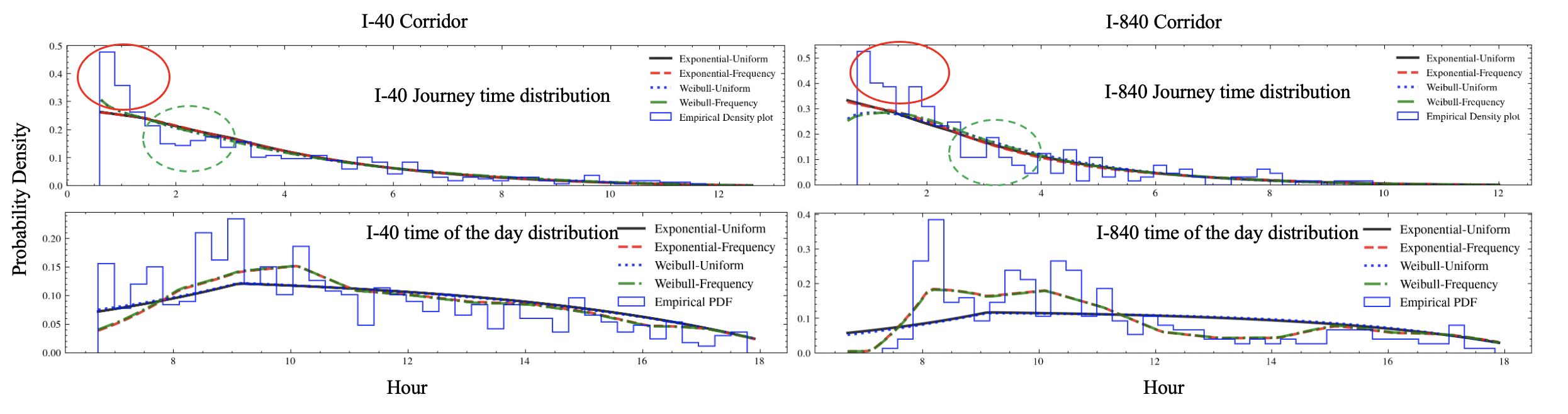}
\centering
\caption{Marginal probability density distribution over the travel time and the arrival time at upstream station}
\label{fig:dist_summ}
\end{figure}

Although not all zones of interest (Zone 1-3) are observable, one can still draw marginal histograms over journey time and traveling time within the observable zone (Zone 2). \textbf{Figure \ref{fig:dist_summ}} compares the empirical distribution against the marginal distribution of the fitted truncated model. For journey time along the I40 corridor, the fitted model underestimates the trucks with journey times under 1.25 hours (see red oval) but overestimates the trucks with journey times between 1.25-3 hours (see green oval with dashed line). Less than expected trucks are re-identified with journey time around 2-3 hours. Assuming 60 minutes are consumed traveling on roads, the remaining 60-120 minutes may not enough for scheduling logitsitcs operations.

\newpage

\begin{figure}[!ht]
\includegraphics[width=\textwidth]{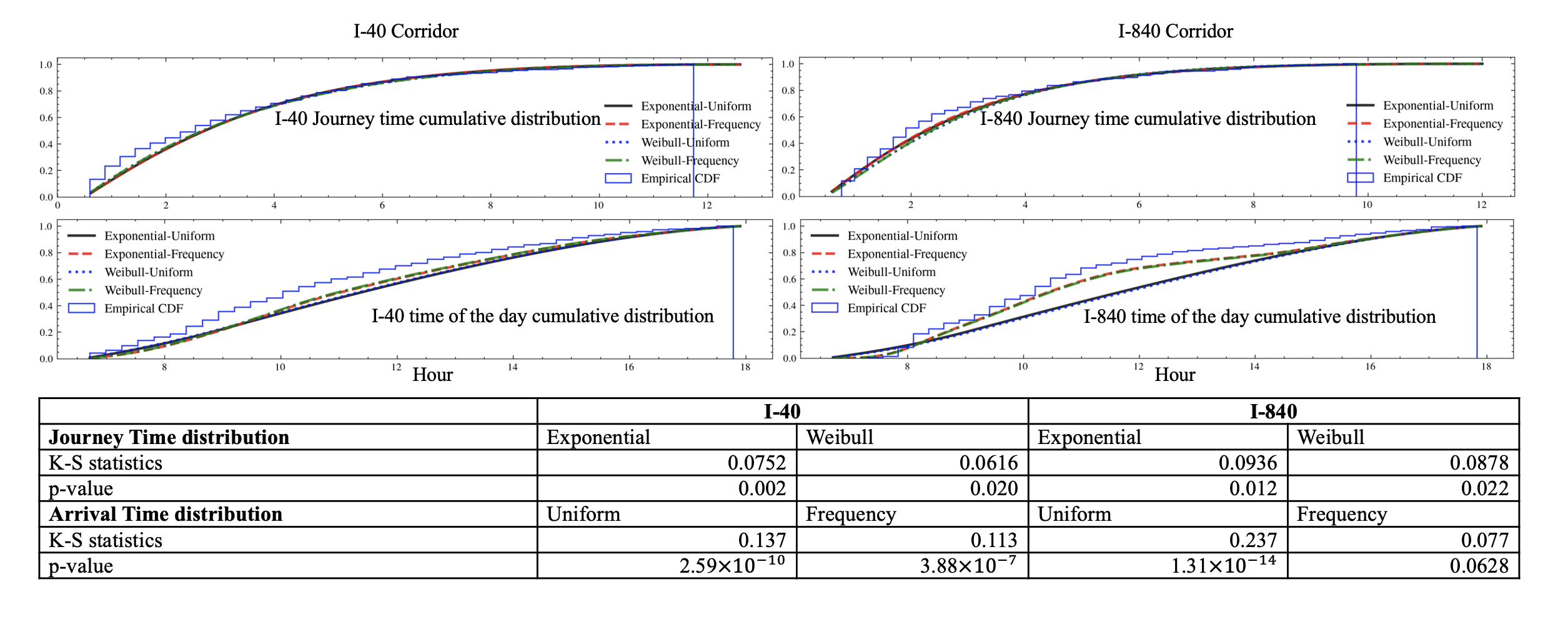}
\centering
\caption{Marginal cumulative probability distribution over the travel time and the arrival time at upstream station}
\label{fig:cdf}
\end{figure}

Besides comparing marginal PDFs, CDFs are also given in \textbf{Figure \ref{fig:cdf}}. Under the significance level of 0.05, most of the null hypotheses are rejected, and it is concluded that the chance to replicate observed data using fitted model is small. Still, consider the small sample size, using one-/two-parameter models are good enough to describe the trend in general without over-fitting the data.

\begin{figure}[!ht]
\includegraphics[width=\textwidth]{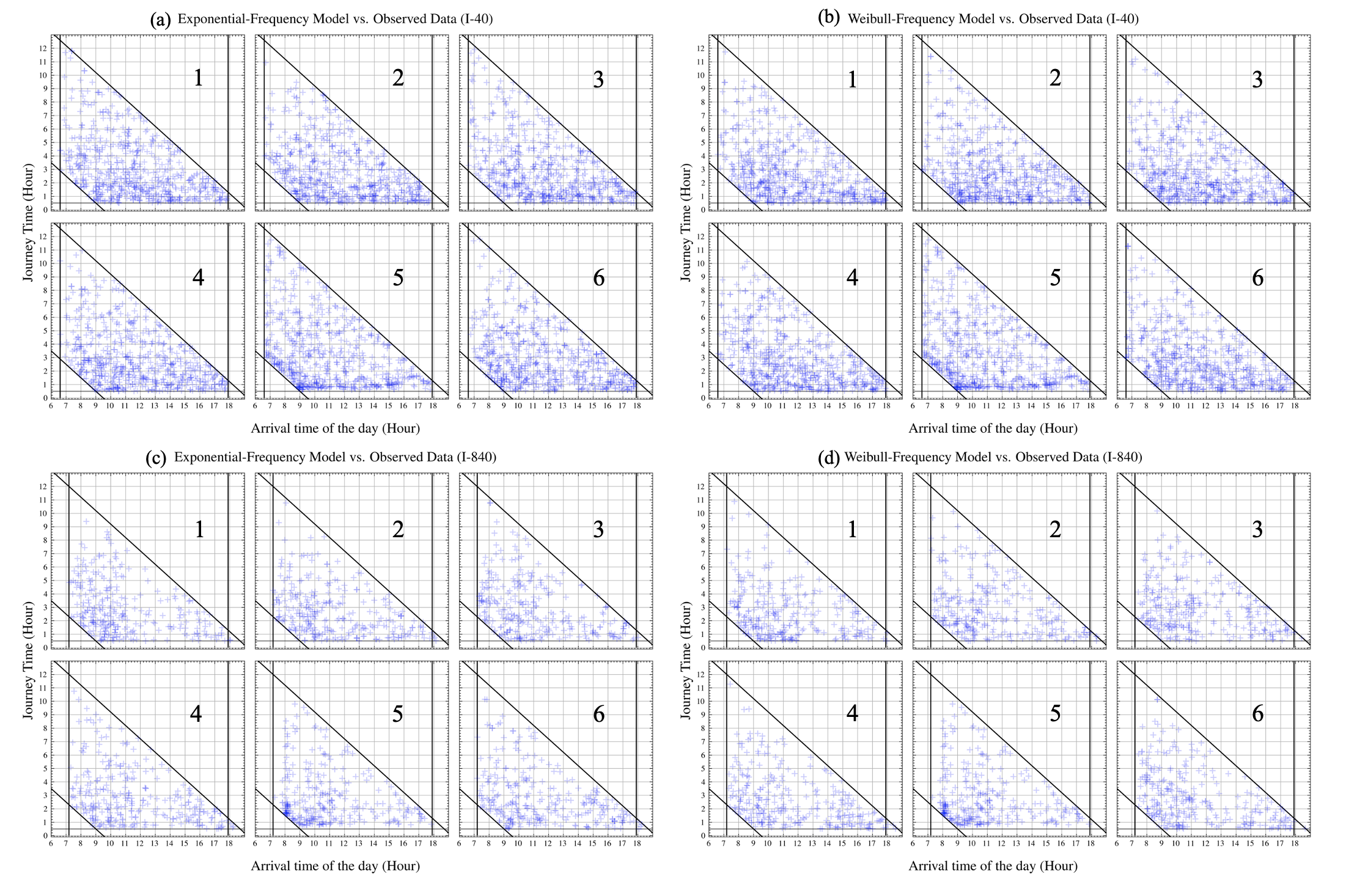}
\centering
\caption{Observed data vs. replicated data using fitted model}
\label{fig:resample}
\end{figure}

Furthermore, Monte Carlo simulations can be applied to generate replicated data-sets given assumed distribution. In the \textbf{Figures \ref{fig:resample}}, the Exponential-Frequency/Weibull-Frequency distributions are re-sampled five times for each route. Among six subplots of each figure, all are replicated data-sets except that sub-figure indexed as 5 as the original observed data. Compared to Exponential-Frequency model, the Weibull-Frequency model is slightly better at imitating the observed data with higher densities among short journey time vehicles. Also, it worth to point out that given fitted models, distribution/counts of the region outside the observable zone can also be speculated.

\newpage
\section{5. Conclusion}
In this paper, we discuss a long-ignored survivorship bias effects in the vehicle re-identification problem where factors such as labor, daylight, legislation, budgets, power, disk memory can all limit the observing time scope. Limited by observable time at data collecting stations, some vehicles cannot be re-identified as they fall out of the "observable zone". Two-point survey, the simplest case of vehicule re-identification survey, is introduced and defined as re-identifying vehicles by matching information collected at two differently located fixed stations. Using two-point survey, the survivorship bias effects and their potential hazards are thoroughly explained using the time-time diagram. To model distribution within the bizarre shape in time-time diagram, two-dimensional truncated parametric models are established where the journey time is modeled as Weibull/Exponential distributions and the upstream arrival time is estimated by Uniform/empirical distributions, giving a combination of four models. MLE is applied to estimate parameters of the models. Fisher Information and the Bootstrap method are further applied to estimate the confidence intervals on top of MLE.

An experiment based on data generated by Monte Carlo simulation is designed to test the validity of proposed method. After that, four formulated models are applied to a real two-point survey investigating trucks' journey time over the I-40/I-840 travel corridors near Nashville, TN. Results show that these simple one/two parameter models are useful in describing overall journey time trends. Compared to the specific distribution assumed describing the journey time, the use of truncated distribution to model the bizarrely shaped observational zone is more important to deal with the survivorship bias effects.

The limited observable scope is not a fantasy but a realistic challenge in understanding truck's logistics behavior. Due to labor shortages, many weigh stations don't operate in the late/early hours, generating a limited observable zone with biased samples observed. Researchers should keep in mind of the potential hazards of the survivorship bias effects and consider using time-time diagram plus truncated distribution to correctly model the problem.

The proposed method can be extended in many directions. With more observed data collected and more explanatory variables (e.g., trucks weight, ownership, etc.) gathered, the proposed method can be extended to model the journey time using those variables. By establishing several observing stations at a set of designated locations, the proposed method can be automated to re-identify trucks' journey time between observing points along the skeleton highways. Using the cordon line approach, insights from the metropolitan level to the state level can be easily gained considering the survivorship bias effects. Besides transportation, the proposed method may also be useful in other engineering fields. For example, the proposed method can help predict the number of end-of-life (EOL) products in near future by considering both the product's lifespan and the number of items produced over time.

\section{6. Acknowledgement}
This research is supported by the State Planning and Research (SPR) Program by Tennessee DOT. We thank Dr. Hyeonsup Lim, Dr. Stephanie Hargrove, Dr. Zhihua Zhang ,and Dr. Yuandong Liu for recording and processing the data. We also thank Dr. Qihao Zhang from Iowa State University for proofreading the Mathematical derivation steps. Thank Lisa Lee's long-enduring efforts in proofreading.

\newpage
\bibliographystyle{trb}
\bibliography{references.bib}

\end{document}